\documentclass[twocolumn,showpacs,preprintnumbers,amsmath,amssymb,superscriptaddress]{revtex4}

\usepackage{graphicx}
\usepackage{dcolumn}
\usepackage{bm}

\begin{document}

\title{Fringe spacing and phase of interfering matter waves}

\author{O. Vainio}
\affiliation{School of Physical Sciences, University of
Queensland, St Lucia, Qld 4072, Australia} \affiliation{Department
of Physics, University of Turku, FIN-20014, Turku, Finland}
\author{C. J. Vale}
\email{vale@physics.uq.edu.au}
\author{M. J. Davis}
\author{N. R. Heckenberg}
\author{H. Rubinsztein-Dunlop}
\affiliation{School of Physical Sciences, University of
Queensland, St Lucia, Qld 4072, Australia}


\date{\today}

\begin{abstract}
We experimentally investigate the outcoupling of atoms from
Bose-Einstein condensates using two radio-frequency (rf) fields in
the presence of gravity. We show that the fringe separation in the
resulting interference pattern derives entirely from the energy
difference between the two rf fields and not the gravitational
potential difference.  We subsequently demonstrate how the phase
and polarisation of the rf radiation directly control the phase of
the matter wave interference and provide a semi-classical
interpretation of the results.
\end{abstract}

\maketitle

Ever since the first realisations of Bose-Einstein condensates
(BECs) in dilute atomic gases, their coherence properties have
been the subject of much investigation. The first clear
demonstration that BECs possess long range phase coherence was
through the interference of two spatially separated condensates
\cite{andrews97}.  Since then, other experiments have studied the
coherence properties using atom laser output from an array of
tunnel coupled condensates in an optical standing wave
\cite{anderson98}, Bragg spectroscopy \cite{stenger98}, density
fluctuations \cite{dettmer01} and intereferometry
\cite{hellweg03,hugbart05}. An elegant scheme to probe condensate
coherence, based on interfering atom laser beams, was reported by
Bloch {\it et al.}  \cite{bloch00,esslinger00}.  This used two
radio frequency (rf) fields to outcouple atoms from different
locations within a condensate.  A high contrast matter wave
interference pattern was observed at temperatures well below the
BEC transition temperature, confirming the phase coherence of the
condensate.  A numerical model of two outcoupled modes agreed with
the experimental observations \cite{schneider00}. More recently,
the atom-by-atom build up of a matter wave interference pattern
has been observed using single atom detection \cite{bourdel05}. To
date, experiments have primarily focussed on the visibility of the
interference patterns. In this paper, we describe experiments
which address the fringe spacing, phase and nature of the
interference.

Outcoupling atoms from Bose-Einstein condensates with rf fields
has been used extensively to produce beams of atoms, generally
referred to as ``atom lasers'' \cite{bloch99,lecoq01,robins04}.
The rf radiation drives resonant (stimulated) transitions from a
trapped Zeeman sublevel to an untrapped state in which the atom
falls under gravity. Outcoupling occurs at locations where the
total energy difference between the trapped and untrapped states
is equal to $\hbar \omega_{rf}$.  This is usually determined by
the Zeeman potential so that the resonant condition may be written
\begin{equation} \hbar \omega_{rf} = \mu_B g_F |B(\textbf{r})|
\label{eqn:resonance}
\end{equation}
where $\mu_B$ is the Bohr magneton, $g_F$ is the Land\'{e}
$g$-factor and $B(\textbf{r})$ is the magnetic field.

Consider a condensate trapped in a cigar-shaped magnetic potential
of the form $U(\textbf{r}) = m \omega_z^2 (\kappa^2 x^2 + y^2 +
z^2)/2$ where $m$ is the mass of the atom, $\omega_z = \omega_y$
is the trapping frequency in the tight directions of the trap and
$\kappa = \omega_x / \omega_z$.  Atoms can be outcoupled from the
surface of an ellipsoid of the magnetic equipotential which
satisfies the resonance condition (\ref{eqn:resonance}).  However,
gravity will cause a displacement of the minimum of the total
potential from the magnetic field minimum. This gravitational sag
means a harmonically trapped condensate will be displaced from the
magnetic field minimum by a distance, $z_0 = -g/\omega_z^2$, where
$\omega_z$ is the trapping frequency in the direction of gravity.
This displacement is typically greater than the size of the
condensate, so that the ellipsoidal equipotential surfaces can be
approximated by planes which intersect the condensate at different
heights, $z$.  In this situation, the dependence on the $x$ and
$y$ coordinates can be neglected for many quantitative purposes
and only the $z$ dimension need be considered.

In previous work \cite{bloch00,esslinger00}, two rf fields of
frequencies, $\omega_1$ and $\omega_2$, were used to outcouple
atoms from a condensate. The two spatially separated resonances
were interpreted as creating two slits from which atoms were
extracted from the condensate. The outcoupled atoms formed two
matter waves which interfered, in close analogy with a Young's
double slit experiment. The visibility of the interference pattern
provided a measure of the first order phase coherence of the
condensate.

The outcoupling points, $z_1$ and $z_2$, used in
\cite{bloch00,esslinger00}, were chosen to be centred around the
middle of the condensate, located at $z_0$, the minimum of the
combined magnetic (harmonic) and gravitational (linear) potential.
Under this condition, the gravitational energy difference between
the two outcoupling points, determined by the slit separation,
$\Delta z = z_1 - z_2$, is exactly equal to the difference in
energy between the two applied rf fields. This can easily be seen
from the derivative of the magnetic potential, where $\Delta E
\approx m \omega_z^2 z \Delta z$. At the central position, $z_0$,
we find $\Delta E = \hbar (\omega_1 - \omega_2) = mg \Delta z$.

\begin{figure}
\includegraphics[width=3.0in]{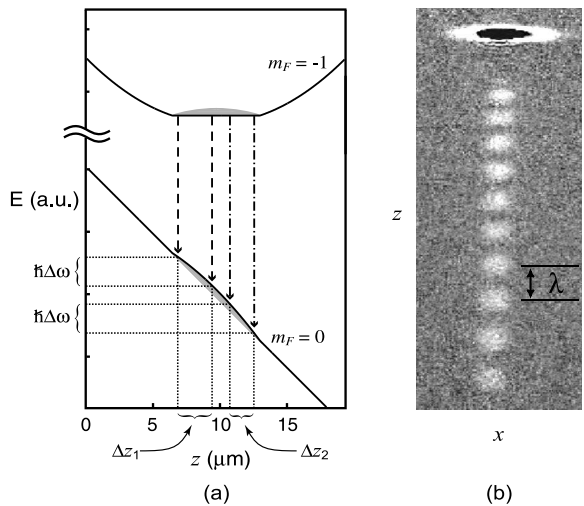}
\caption{(a) Total energy (solid lines) of atoms in the trapped,
$m_F = -1$ and untrapped $m_F = 0$ states, for the parameters used
in our experiments. The dashed and dash-dotted lines represent two
pairs of rf fields, with equal $\Delta \omega$, used to outcouple
atoms from the BEC at different locations. The gravitational
energy difference $m g \Delta z$ may vary for a fixed $\Delta
\omega$, however, when the interaction energy (shaded) of atoms in
the untrapped state is included, the total energy difference
between the two outcoupled matter waves is always equal to $\hbar
\Delta \omega$. (b) The measured outcoupled matter wave beams
display an interference pattern with a constant fringe spacing
$\lambda(z)$ given by equation \ref{eqn:lambda} for a fixed
$\Delta \omega$ (in this case equal to $2 \pi \times
1000$\,s$^{-1}$), independent of $\Delta z$.} \label{fig:espectra}
\end{figure}

However, this result is only true when $\bar{z} = z_0$, (where
$\bar{z} = (z_1 + z_2)/2$, is the distance from the magnetic field
minimum to the centre of the two slits). If the two resonant
points are not symmetrically located about the centre of the trap,
the $z^2$ dependence of the magnetic potential means that the slit
separation for a fixed $\Delta \omega = \omega_1 - \omega_2$
varies inversely with $\bar{z}$. Thus the gravitational energy
difference between the two resonance positions, is not necessarily
equal to $\hbar \Delta \omega$ and $\Delta z$ may change
significantly across the width of a condensate. For a harmonic
potential, the slit separation is approximately
\begin{equation}\label{eqn:deltaz}
   \Delta z \approx \frac{\hbar \Delta \omega}{m \omega_z^2 \bar{z}}.
\end{equation}

In a Young's double slit experiment, the fringe spacing,
$\lambda$, of the interference pattern is proportional to $\Delta
z^{-1}$. However, in dual rf outcoupling experiments with a fixed
$\Delta \omega$, $\lambda$ is independent of $\Delta z$. While the
gravitational energy difference, $m g \Delta z$, between the two
resonant locations can change, this is not the only energy to
consider. The fringe spacing of the interference pattern depends
on the total energy difference between the two indistinguishable
outcoupling paths and must always equal $\hbar \Delta \omega$ to
satisfy energy conservation. This is a general result which we
discuss below, for a BEC in the Thomas-Fermi (TF) regime.

Consider the specific case of an $F = 1$ $^{87}$Rb TF condensate.
The total energy of the trapped state $| F=1, m_F=-1 \rangle$
consists of the sum of its magnetic, gravitational and mean field
energies. The untrapped (outcoupled) state $| 1, 0 \rangle$
experiences negligible magnetic potential, but, while still within
the condensate, experiences both the mean field and gravitational
potentials. In 1D, the total energy of a particular substate can
be written as
\begin{equation}\label{eqn:Etot}
   E_{m_F}(z) = -m_F ( \frac{1}{2} m \omega_z^2 z^2 + \mu_B g_F B_0 )
   - m g z + g_{\rm 1d} |\psi (z)|^2
\end{equation}
where $B_0$ is the magnetic field at the minimum of the trap,
$g_{\rm 1d}$ is the 1D effective interaction strength (assumed to
be the same for all $m_F$, which is approximately true but not an
essential point in this discussion) and $|\psi (z)|^2 = \sum_{m_F}
|\psi_{m_F} (z)|^2$ is the total atomic density.  The energies of
the trapped $| 1, -1 \rangle$ state and the untrapped $| 1, 0
\rangle$ state are plotted (solid lines) in Fig.
\ref{fig:espectra}(a) for the parameters used in our experiments.
The shaded regions indicate the mean field contribution to the
total energy.  Also shown are two pairs of rf fields, dashed and
dash-dotted lines, with the same $\Delta \omega$, chosen to lie
within the width of the condensate. As the two pairs are centred
around different $\bar{z}$, the resulting $\Delta z$ for each pair
is different, but the total energy difference is $\hbar \Delta
\omega$.

In the TF limit the interaction energy between the condensate and
the outcoupled state exactly compensates for the difference in
gravitational potential at different slit locations.  The density
profile, $|\psi (z)|^2$, mirrors the shape of the magnetic
trapping potential so that the energy splitting between the two
states is always given by the difference in their magnetic
potentials. Additionally, the energy of trapped atoms within the
condensate is independent of $z$ so that only the final energies
on the $m_F = 0$ curve determine the energy difference between the
two outcoupled beams.

Having established that the fringe spacing, $\lambda$, depends
only on $\Delta \omega$, it can easily be shown that
\begin{equation}\label{eqn:lambda}
   \lambda(z) = \frac{\sqrt{2 g (z - z_0)}}{\Delta f},
\end{equation}
where $\Delta f = \Delta \omega / 2 \pi$.  $\lambda$ varies with
$z$ because the outcoupled atoms accelerate in the $z$-direction
under gravity, as can be seen in Fig. \ref{fig:espectra}(b).

We have performed a range of experiments to verify this for
several values of $\Delta \omega$ with the resonant points centred
around various $z$ positions. Our experimental procedure for
producing condensates has been described elsewhere \cite{vale04}
and was used here with only slight modifications. An atom chip is
used to produce near pure condensates containing $2 \times 10^5$
$^{87}$Rb atoms in the $| 1, -1 \rangle$ ground state.  Our chip
design facilitates the production of relatively large condensates
in highly stable trapping fields.  The final trapping frequencies
are 160\,Hz in the tight direction and 6.7\,Hz in the weak
direction. The elongated geometry of the trap means we must cool
well below the 3D critical temperature $T_c$ to produce fully
phase coherent condensates (typically $T_\phi$ for our parameters
is less than $T_c/2$ \cite{petrov01}). Outcoupling is induced by
turning on two rf fields of the same amplitude, with frequencies
$\omega_1$ and $\omega_2$ tuned to be resonant with atoms in the
condensate, and Rabi frequencies, $\Omega = \mu_B g_F |B| / 2
\hbar$, of 50\,Hz for each rf source.  After outcoupling for
10\,ms, the trap is left on for a further 3\,ms before being
turned off abruptly.  An absorption image is taken after 5.3\,ms
of free expansion.

We first checked the reproducibility of the fringe spacing for a
fixed $\Delta \omega$ of $2 \pi \times 1000$\,s$^{-1}$ as $\Delta
z$ was varied over the range 390\,nm to 560\,nm by varying
$\bar{z}$. While the visibility of the interference pattern
decreased near the edges of the condensate, the wavelength was
consistent with the value predicted by equation (\ref{eqn:lambda})
to within 1$\%$.  Experimental values of $\lambda$ were determined
by taking an image of the outcoupled atoms, similar to that shown
in Fig. \ref{fig:espectra}(b), converting the $z$ spatial axis
into a time axis through the relation $t = \sqrt{2(z-z_0)/g}$,
integrating the output over $x$ and fitting a $\cos^2$ function to
the data. The uncertainty in $\lambda$ is determined by the
uncertainty in the fitted frequency of the $\cos^2$ function.
Variations at the level of 1$\%$ are within our experimental
uncertainties and not significant when compared to what would be
expected if $\lambda$ was determined by $\Delta z^{-1}$.  This
would lead to a variation of more than $30\%$ across the range of
values we measured.

A semi-classical interpretation of these dual rf outcoupling
experiments, based on the interference of the applied rf fields,
may also be used to understand these experiments. Defining the
mean frequency, $\bar{\omega} = (\omega_1 + \omega_2)/2$, and the
beat frequency, $\delta = (\omega_1 - \omega_2)/2$, and recalling
the standard trigonometric identity
\begin{equation}\label{eqn:2rfs}
   \sin \omega_1 t + \sin \omega_2 t = 2 \sin \bar{\omega} t
   \cos \delta t
\end{equation}
we see that the sum of two oscillating fields is equivalent to an
amplitude modulated carrier wave.  In our case, the carrier
frequency, $\bar{\omega}$, is typically three orders of magnitude
higher than the beat frequency, $\delta$. Adding a phase, $\phi$,
to one of the rf fields shifts the phase of both the carrier and
beating terms by half this amount.

\begin{figure}\label{fig:alrfphaseb}
\includegraphics[width=3.2in]{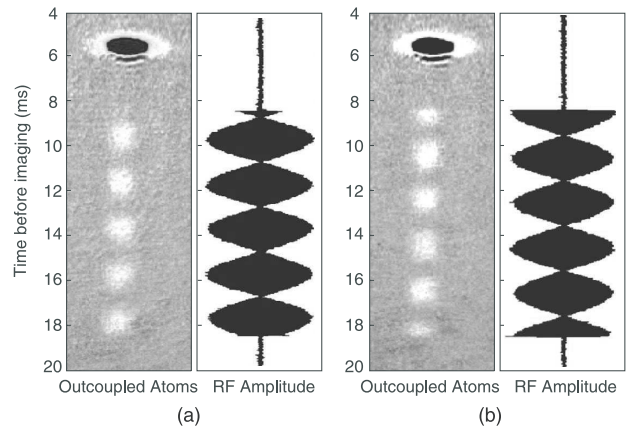}
\caption{The phase of the interfering matter wave beams is
determined by the phase of the beating rf fields used in to drive
the outcoupling.  (a) and (b) represent different runs of the
experiment under identical conditions apart from different phases
of the applied rf field.  On the left are absorption images of the
condensate (top) and outcoupled atoms and on the right is the beat
note of the corresponding rf used to drive the outcoupling,
measured on an oscilloscope.  The vertical axis indicates the time
before the image was taken and was obtained for the atom images
through the relationship, $t = \sqrt{2(z-z_0)/g}$}
\label{fig:alrfphase}
\end{figure}

We may now consider the condensate interacting with a single rf
field at the carrier frequency, which is amplitude modulated in
time. The number of atoms outcoupled is proportional to the Rabi
frequency squared (ie. proportional to the amplitude squared of
the rf field at time $t$) and is modulated at $2\delta = \Delta
\omega$. Once outcoupled, the atoms fall under gravity and,
provided they have a low spread of initial momenta, the outcoupled
density will be modulated in time.

To demonstrate this, a sequence of dual rf outcoupling experiments
was performed using fixed values $\omega_1$ and $\omega_2$
($\Delta \omega = 2 \pi \times 500\,s^{-1}$) but varying the
relative phase of the two rf fields. This has the effect of
shifting the phase of the rf beat note. All other experimental
parameters were kept fixed. Two examples of the data obtained are
shown in Fig. \ref{fig:alrfphase}. Absorption images of the
outcoupled atoms appear on the left and the beat note of the
corresponding rf fields used for outcoupling (measured on an
oscilloscope) are shown on the right. The $z$ axis of the
absorption images has been rescaled by $\sqrt{2(z-z_0)/g}$ to
linearise the time axis for ease of comparison with the rf. It is
clear from these images that the outcoupled atoms correspond to
the largest amplitude of the rf beat note.

In order to quantify this, we have analysed a series of similar
data in which the relative phase of the rf fields was allowed to
vary randomly over the range 0 to 2$\pi$.  The fitting procedure
described earlier was applied to all of the absorption images to
determine the phase of the outcoupled beam.  A similar fit was
applied to the square of the measured rf beat note and the two
phases are plotted against each other (filled sircles) in Fig.
\ref{fig:phases}. The dashed line through this data is a plot of
$y = x$.  The phase of the modulated atom beam matches very well
the phase of the applied rf field.

In these experiments both rf fields were provided by passing the
two rf currents through the same coil.  This means the rf field in
the vicinity of the BEC was linearly polarised, perpendicular to
the quantisation axis $x$.  We have also performed experiments
using separate, orthogonally mounted coils where each rf current
was sent through a different coil. The rf interference is no
longer linearly polarised but rather a field whose polarisation
varies from vertical linear, to left hand circular, to horizontal
linear, to right hand circular and back to vertical linear in a
single beat period ($T = 1/\Delta \omega$). This is analogous to
the optical field used in lin $\perp$ lin sub-Doppler (Sisyphus)
laser cooling \cite{dalibard89} but the field is periodic in time
rather than space.

The outcoupling transition from $| 1, -1 \rangle$ to $| 1, 0
\rangle$ requires $+\hbar$ of angular momentum, and can only be
driven by $\sigma^+$ radiation.  When the rf is derived from a
single coil, it is easy to see that the maximum amplitude of
$\sigma^+$ radiation occurs when the two rf fields are in phase.
The linearly polarised field may be decomposed into two counter
rotating circular fields and it is the $\sigma^+$ component of
this which couples to the atom.  With perpendicularly oriented
coils however, the maximum amplitude of the $\sigma^+$ field
occurs during the circular polarised phase of the beat note. This
happens when the rf fields are $\pi/2$ out of phase.

Also shown in Fig. \ref{fig:phases} (open squares) is a plot of
the phase of the atom laser versus the phase of the beating rf
fields performed with (near) perpendicularly oriented coils. The
atomic output is phase shifted by approximately $\pi/2$ from the
rf as expected. The slight mismatch between the measured shift of
$0.55 \pi$ and the expected shift of $0.5 \pi$ was due to
imperfect alignment of the rf coils (precise perpendicular
alignment would have impeded optical access in our setup).  For
coils mounted antiparallel maximum outcoupling would occur when
the two rf fields are $\pi$ out of phase.

\begin{figure}\label{fig:phases}
\includegraphics[width=3.2in]{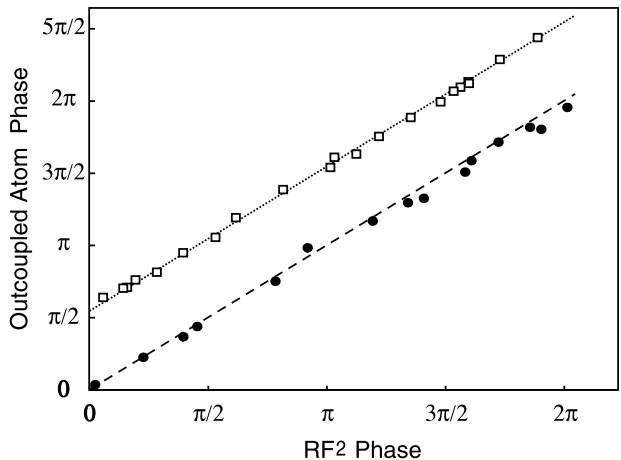}
\caption{Plot of the fitted phase of the modulated atom beam
against the phase of the beating rf field.  Filled circles
represent experimentally measured data using a single coil and the
dashed line is a plot of $y = x$.  Open squares are experimentally
measured phases when the rf is produced by two separate (near)
orthogonally mounted coils and the dotted line is a plot of $y = x
+ 0.55 \pi$.} \label{fig:phases}
\end{figure}

In conclusion, we have studied the origins of the fringe spacing
and phase of matter wave interference patterns, produced by
outcoupling atoms from a BEC with two rf fields. We have shown
that the energy difference between the two rf fields determines
the spacing of the interference pattern, not the gravitational
potential difference determined from the classical slit
separation. Semi-classical arguments based on interfering rf
fields correctly predict the experimental observations. These also
show how the phase and polarisation of the rf field determine the
phase of the observed matter wave interference pattern. This
extends previous work which looked at the fringe visibility for
the specific case where $\hbar \Delta \omega = m g \Delta z$
\cite{bloch00}. Our findings do not contradict the phase coherence
studies reported in ref. \cite{bloch00}. Indeed, any random phase
gradients within the BEC would lead to random initial velocities
that would degrade the observed interference patterns. Finally, we
note that we have also performed experiments with cold thermal
atoms and see (as in \cite{bloch00}) that the visibility of
interference pattern diminishes, due to the thermal spread of
velocities in the trapped gas.

We acknowledge valuable discussions with Craig Savage, Ashton
Bradley and Murray Olsen and technical assistance from Evan Jones.
O. V. acknowledges financial support from the Jenny and Antti
Wihuri Foundation and the Academy of Finland (project 206108).
This work was supported by the Australian Research Council.

\end{document}